\documentclass[prl,aps,twocolumn,showpacs,superscriptaddress,nofootinbib]{revtex4}

\usepackage{graphicx}
\usepackage{amssymb}
\usepackage{amsmath}
\usepackage{color}

\newcommand{\ip}[2]{\langle #1|#2 \rangle}
\newcommand{\ket}[1]{|#1 \rangle}
\newcommand{\bra}[1]{\langle #1 |}

\newcommand{\tr}{\mathrm{Tr}}

\newcommand{\Lx}{\protect\overleftarrow{X}}
\newcommand{\Rx}{\protect\overrightarrow{X}}
\newcommand{\lx}{\protect\overleftarrow{x}}
\newcommand{\rx}{\protect\overrightarrow{x}}

\begin{document}

\title{Occam's Quantum Razor: How Quantum Mechanics can reduce the complexity of classical models}

\author{Mile Gu}
\affiliation {Center for Quantum Technology, National University of Singapore, Republic of Singapore}

\author{Karoline Wiesner}
\affiliation {School of Mathematics, Centre for Complexity Sciences, University of Bristol, Bristol BS8 1TW, United Kingdom}

\author{Elisabeth Rieper}
\affiliation {Center for Quantum Technology, National University of Singapore, Republic of Singapore}

\author{Vlatko Vedral}
\affiliation{Atomic and Laser Physics, Clarendon Laboratory,
University of Oxford, Parks Road, Oxford OX13PU, United Kingdom}
\affiliation{Department of Physics, National University of Singapore,
Republic of Singapore}

\date{\today}

\begin{abstract}
Mathematical models are an essential component of quantitative science. They generate predictions about the future, based on information available in the present. In the spirit of Occam's razor, simpler is better; should two models make identical predictions, the one that requires less input is preferred. Yet, for almost all stochastic processes, even the provably optimal classical models waste information. The amount of input information they demand exceeds the amount of predictive information they output. We systematically construct quantum models that break this classical bound, and show that the system of minimal entropy that simulates such processes must necessarily feature quantum dynamics. This indicates that many observed phenomena could be significantly simpler than classically possible should quantum effects be involved.\end{abstract}


\maketitle

\section{Introduction}
Occam's razor, the principle that `plurality is not to be posited without necessity', is an important heuristic that guides the development of theoretical models in quantitative science. In the words of Isaac Newton,``We are to admit no more causes of natural things than such as are both true and sufficient to explain their appearances." Take for example application of Newton's laws on an apple in free fall. The future trajectory of the apple is entirely determined by a second order differential equation, that requires only its current location and velocity as input. We can certainly construct alternative models that predict identical behavior, that demand the apple color, or its entire past trajectory as input. Such theories, however, are dismissed by Occam's razor, since they demand input information that is either unnecessary or redundant.

Generally, a mathematical model of a system of interest is an algorithmic abstraction of its observable output. Envision that the given system is encased within a black box, such that we observe only its output. Within a second box resides a computer that executes a model of this system with appropriate input. For the model to be accurate, we expect these boxes to be operationally indistinguishable; their output is statistically equivalent, such that no external observer can differentiate which box contains the original system.

\begin{figure*}
	\centering
		 \includegraphics[width=0.75\textwidth]{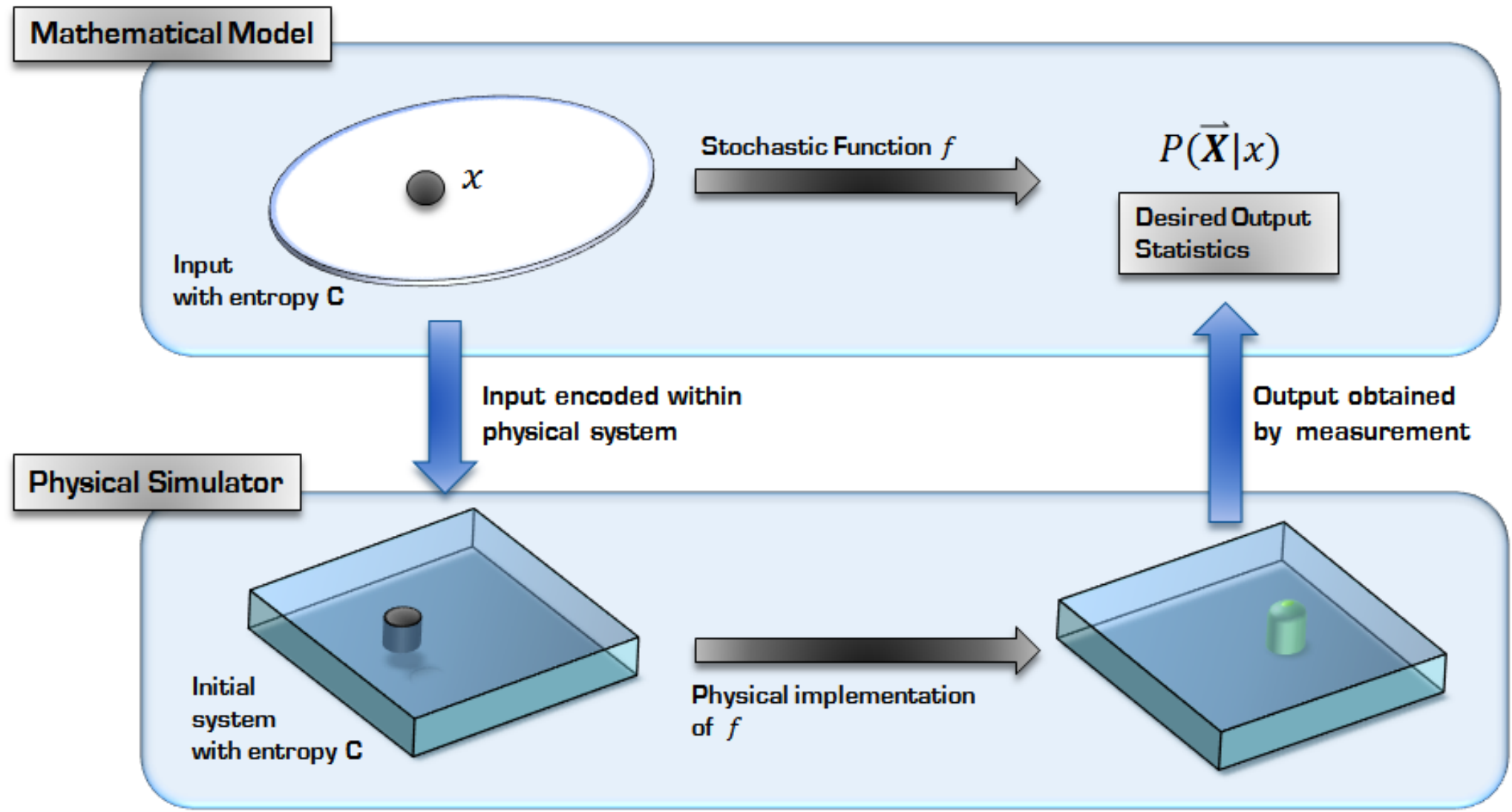}
	\caption{\label{fig:figure_model}\textbf{The Relationship between models and simulators.} A mathematical model is defined by a stochastic function $f$ that maps relevant data from the present,`$x$', to desired output statistics that coincides with the process it seeks to model. To implement this model, we must realize it within some physical simulator. To do this, we (a) encode '$x$' within a suitable physical system, (b) evolve the system according to a physical implementation of $f$ and (c) retrieve the predictions of model by appropriate measurement. On the other hand, given a simulator with entropy $C$ that outputs statistically identical predictions, we can always construct a corresponding mathematical model that takes the initial state of this system as input. Thus the input entropy of a model and the initial entropy of its corresponding simulator coincide (this is also a lower bound on the amount of information the simulator must store). In this article, we regard both models and simulators as algorithms that map input states to desired output statistics, with implicit understanding that the two terms are interchangeable. The former emphasizes the mathematical nature of these algorithms, while the latter their physical realization.}
\end{figure*}

There are numerous distinct models for any given system. Consider a system of interest consisting of two binary switches. At each time-step, the system emits a $0$ or $1$ depending on whether the state of the two switches coincides, and one of the two switches is chosen at random and flipped. The obvious model that simulates this system keeps track of both switches, and thus requires an input of entropy $2$. Yet, the output is simply a sequence of alternating $0$s and $1$s, and can thus be modeled knowing only the value of the previous emission. Occam's razor stipulates that this alternative is more efficient and thus superior; it demands only an input of entropy $1$ (i.e., a single bit), when the original model required two. This motivates a direct interpretation of Occam's razor; the optimal model of a particular behavior is the one whose input is of minimal entropy. Indeed, this interpretation has been already adopted as a principle of computational mechanics~\cite{crutch94,ray04}.

Efficient mathematical models carry operational consequence. The practical application of a model necessitates its physical realization within a corresponding simulator (Fig.~\ref{fig:figure_model}). Therefore, should a model demand an input of entropy $C$, its physical realization must contain the capacity to store that information.  The construction of simpler mathematical models for a given process allows potential construction of simulators with reduced information storage requirements. Thus we can directly infer the minimal complexity of an observed process once we know its simplest model. If a process exhibits observed statistics that require an input of entropy $C$ to model, then whatever the underlying mechanics of the observed process, we require a system of entropy $C$ to simulate its future statistics.

These observations motivate \emph{maximally efficient} models; models that generate desired statistical behavior, while requiring minimal input information. In this article, we show that even when such behavior aligns with simple stochastic processes, such models are almost always quantum. For any given stochastic process, we outline its provably simplest classical model, We show that unless improvement over this optimal classical model violates the second law of thermodynamics, our construction and a superior quantum model and its corresponding simulator can \emph{always} be constructed.

\section{Results}

\textbf{Framework and tools.} We can characterize the observable behavior of any dynamical process by a joint probability distribution $P(\Lx,\Rx)$, where $\Lx$ and $\Rx$ are random variables that govern the system's observed behavior respectively, in the past and the future. Each particular realization of the process has a particular past $\lx$, with probability $P(\Lx = \lx)$. Should there exists a model for this behavior with an input of entropy $C$, then we may compress $\lx$ within a system $\mathcal{S}$ of entropy $C$, such that systematic actions on $\mathcal{S}$ generates random variables whose statistics obey $P(\Rx|\Lx = \lx)$.

We seek the \emph{maximally efficient} model, such that $C$ is minimized. Since the past contains exactly $\mathbf{E} = I(\Lx:\Rx)$ (the mutual information between past and future) about the future, the model must require an input of entropy at least $\mathbf{E}$ (this remains true for quantum systems~\cite{Holevo73a}). On the other hand, there appears no obvious reason a model should require anything more. We say that the resulting model, where $C = \mathbf{E} = I(\Lx:\Rx)$, is \emph{ideal}. It turns out that for many systems such models do not exist.

Consider a dynamical system observed at discrete times $t \in \mathbb{Z}$, with possible discrete outcomes $x_t \in \Sigma$ dictated by random variables $X_t$. Such a system can be modeled by a stochastic process\cite{Doob53a}, where each realization is specified by a sequence of past outcomes $\lx = \ldots x_{-3}x_{-2}x_{-1}$, and exhibits a particular future $\rx = x_0x_1x_2\ldots$ with probability $P(\Rx =\rx|\Lx = \lx)$. Here, $\mathbf{E} = I(\Lx:\Rx)$, referred to as \emph{excess entropy}\cite{crutchfield03,Grassberger86}, is a quantity of relevance in diverse disciplines ranging from spin systems~\cite{crutchfield97a} to measures of brain complexity\cite{Tononi1994}. How can we construct the simplest simulator of such behavior, preferably with input entropy of no more than $\mathbf{E}$?

The brute force approach is to create an algorithm that samples from $P(\Rx|\Lx = \lx)$ given complete knowledge of $\lx$. Such a construction accepts $\lx$ directly as input, resulting in the required entropy of $C = H(\Lx)$, where $H(\Lx)$ denotes the Shannon entropy of the complete past. This is wasteful. Consider the output statistics resulting from a sequence of coin flips, such that $P(\Lx,\Rx)$ is the uniform distribution over all binary strings. $\mathbf{E}$ equals $0$ and yet $C$ is infinite. It should not require infinite memory to mimic a single coin, better approaches exist.

\textbf{Simplest classical models.} $\epsilon$-machines are the provably optimal classical solution\cite{crutch89,crutch01}. They rest on the rationale that to exhibit desired future statistics, a system needs not distinguish differing pasts, $\lx$ and $\lx'$, if their future statistics coincide. This motivates the equivalence relation, $\sim$, on the set of all past output histories, such that $\lx \sim \lx'$ iff $P(\Rx|\lx) = P(\Rx|\lx')$. To sample from $P(\Rx|\lx)$ for a particular $\lx$, a $\epsilon$-machine need not store $\lx$, only which equivalence class, $\epsilon(\lx) \equiv \{\lx': \lx \sim \lx'\}$, $\lx$ belongs to. Each equivalence classes is referred to as a causal state.

For any stochastic process $P(\Lx,\Rx)$ with emission alphabet $\Sigma$, we may deduce its causal states $\{S_i\}_{i=1}^N$ that form the state space of its corresponding $\epsilon$-machine. At each time step $t$, the machine operates according to a set of transition probabilities $T_{j,k}^{(r)}$; the probability that the machine will output $x_t = r \in \Sigma$, and transition to $S_k$ given that it is in state $S_j$. The resulting $\epsilon$-machine, when initially set to state $\epsilon(\lx)$, generates a sequence $\rx$ according to probability distribution $P(\Rx|\Lx = \lx)$ as it iterates through these transitions. The resulting $\epsilon$-machine thus has internal entropy
\begin{equation}
C = H(\mathbf{S}) = - \sum_{j
 \in \mathcal{S}} p_j \log p_j \equiv C_\mu
\end{equation}
where $\mathbf{S}$ is the random variable that governs $S_j = \epsilon(\lx)$ and $p_j$ is the probability that $\epsilon(\lx) = S_j$.

The provable optimality of $\epsilon$-machines among all classical models motivates $C_\mu$ as an intrinsic property of a given stochastic process, rather than just a property of $\epsilon$-machines. Referred to in literature as the statistical complexity~\cite{crutch01,crutch09}, its interpretation as the minimal amount of information storage required to simulate such a given process has been applied to quantify self-organization~\cite{Shaliz93}, the onset of chaos~\cite{crutch89} and complexity of protein configuration space~\cite{Chun-BiuLi08}. Such interpretations, however, implicitly assume that classical models are optimal. Should a quantum simulator be capable of exhibiting the same output statistics with reduced entropy, this fundamental interpretation of $C_\mu$ may require review.

\textbf{Classical models are not ideal.} There is certainly room for improvement. For many stochastic processes, $C_\mu$ is strictly greater than $\mathbf{E}$~\cite{crutch09}; the $\epsilon$-machine that models such processes is fundamentally irreversible. Even if the entire future output of such an $\epsilon$-machine was observed, we would still remain uncertain which causal state the machine was initialized in. Some of that information has been erased, and thus, in principle, need never be stored. In this paper we show that for \emph{all} such processes, quantum processing helps; for \emph{any} $\epsilon$-machine such that $C_\mu > \mathbf{E}$, there exists a quantum system, a quantum $\epsilon$-machine with entropy $C_q$, such that $C_\mu > C_q \geq \mathbf{E}$. Therefore, the corresponding model demands an input with entropy no greater than $C_q$.

The key intuition for our construction lies in identifying the cause of irreversibility within classical $\epsilon$-machines, and addressing it within quantum dynamics. An $\epsilon$-machine distinguishes two different causal states provided they have differing future statistics, but makes no distinction based on \emph{how much} these futures differ. Consider two causal states, $S_j$ or $S_k$, that both have potential to emit output $r$ at the next time-step and transition to some coinciding causal state $S_l$. Should this occur, some of the information required to completely distinguish $S_j$ and $S_k$ has been irreversibly lost. We say that $S_j$ and $S_k$ share non-distinct futures. In fact, this is both necessary and sufficient condition for $C_\mu >\mathbf{E}$ (See methods for proof).

\textbf{The irreversibility condition.} Given a stochastic process $P(\Lx,\Rx)$ with excess entropy $\mathbf{E}$ and statistical complexity $C_\mu$. Let its corresponding $\epsilon$-machine have transition probabilities $T_{j,k}^{(r)}$. Then  $C_\mu > \mathbf{E}$ iff there exists a non-zero probability that two different causal states, $S_j$ and $S_k$ will both make a transition to a coinciding causal state $S_l$ upon emission of a coinciding output $r \in \Sigma$, i.e., $T_{j,l}^{(r)}, T_{k,l}^{(r)} \neq 0$. We refer to this as the \emph{irreversibility condition}.

This condition highlights the fundamental limitation of any classical model. In order to generate desired statistics, any classical model must record each binary property $A$ such that $P(\Rx|A = 0) \neq P(\Rx|A = 1)$, regardless of how much these distributions overlap.  In contrast, quantum models are free of such restriction. A quantum system can store causal states as quantum states that are \emph{not} mutually orthogonal. The resulting quantum $\epsilon$-machine differentiates causal states sufficiently to generate correct statistical behavior. Essentially, they save memory by `partially discarding' $A$, and yet retain enough information to recover statistical differences between $P(\Rx|A = 0)$ and $P(\Rx|A = 1)$.


\textbf{Improved quantum models} Given an $\epsilon$-machine with causal states $S_j$ and transition probabilities $T_{j,k}^{(r)}$, we define quantum causal states
\begin{equation}\label{eqn:qstates}
\ket{S_j} = \sum_{k = 1}^N \sum_{r \in \Sigma} \sqrt{T_{jk}^{(r)}}\ket{r}\ket{k},
\end{equation}
where $\ket{r}$ and $\ket{k}$ form orthogonal bases on Hilbert spaces of size $|\Sigma|$ and $|\mathcal{S}|$ respectively. A quantum $\epsilon$-machine accepts a quantum state $\ket{S_j}$ as input in place of $S_j$. Thus, such a system has an internal entropy of\begin{equation}\label{eqn:qc}
C_q = -\mathrm{Tr} \rho \log \rho,
\end{equation}
where $\rho = \sum_j p_j \ket{S_j}\bra{S_j}$. $C_q$ is clearly strictly less than $C_\mu$ provided not all $\ket{S_j}$ are mutually orthogonal~\cite{Nielsen00a}.

This is guaranteed whenever $C_\mu > \mathbf{E}$. The irreversibility condition implies that there exists two causal states, $S_j$ and $S_k$, which will both make a transition to a coinciding causal state $S_l$ upon emission of a coinciding output $r \in \Sigma$, i.e., $T_{j,l}^{(r)}, T_{k,l}^{(r)} \neq 0$. Consequently $\ip{S_j}{S_k} \geq \sqrt{T_{j,l}^{r}T_{k,l}^{r}} > 0$ iff $T_{j,l}^{(r)}, T_{k,l}^{(r)} \neq 0$, and thus $\ket{S_j}$ is not orthogonal with respect to $\bra{S_j}$.

A quantum $\epsilon$-machine initialized in state $\ket{S_j}$ can synthesis black-box behavior which is statistically identical to a classical $\epsilon$-machine initialized in state $S_j$. A simple method is to (i) measure $\ket{S_j}$ in the basis $\ket{r}\ket{k}$, resulting in measurement values $r,k$. (ii) Set $r$ as output $x_0$ and prepare the quantum state $\ket{S_k}$. Repetition of this process generates a sequence of outputs $x_1,x_2,\ldots$ according to the same probability distribution as the original $\epsilon$-machine and hence $P(\Rx|\lx)$. (We note that while the simplicity of the above method makes it easy to understand and amiable to experimental realization, there's room for improvement. The decoding process prepares $S_k$ based of the value of $k$, and thus still requires $C_\mu$ bits of memory. However, there exist more sophisticated protocols without such limitation, such that the entropy of the quantum $\epsilon$-machine remains at $C_q$ at all times. One is detailed in methods). These observations lead to the central result of our paper.

\emph{Theorem:} Consider any stochastic process $P(\Lx,\Rx)$ with excess entropy $\mathbf{E}$, whose optimal classical model has input entropy $C_\mu > \mathbf{E}$. Then we may construct a quantum system that generates identical statistics, with input entropy $C_q < C_\mu$. In addition, the entropy of this system never exceeds $C_q$ while generating these statistics.

There always exists quantum models of greater efficiency than the optimal classical model, unless the optimal classical model is already ideal.

\textbf{A concrete example of simulating perturbed coins.} We briefly highlight these ideas with a concrete example of a perturbed coin. Consider a process $P(\Lx, \Rx)$ realized by a box that contains a single coin. At each time step, the box is perturbed such that the coin flips with probability $0 < p < 1$, and the state of the coin is then observed. This results in a stochastic process, where each $x_t \in \{0,1\}$, governed by random variable $X_t$, represents the result of the observation at time $t$.

For any $p \neq 0.5$, this system has two causal states, corresponding to the two possible states of the coin; the set of pasts ending in $0$, and the set of pasts ending in $1$. We call these $S_0$ and $S_1$. The perturbed coin is its own best classical model, requiring exactly a system of entropy $C_\mu = 1$, namely the coin itself, to generate correct future statistics.

As $p \rightarrow 0.5$, the future statistics of $S_0$ and $S_1$ become increasingly similar. The stronger the perturbation, the less it matters what state the coin was in prior to perturbation. This is reflected by the observation that $\mathbf{E} \rightarrow 0$ (in fact $\mathbf{E} = 1 - H_s(p)$~\cite{crutchfield97a}, where $H_s(p) = - p \log p - (1-p)\log(1-p)$ is the Shannon entropy of a biased coin that outputs head with probability $p$ \cite{Shannon51a}). Thus only $\mathbf{E}/C_\mu = 1 - H_s(p)$ of the information stored is useful, which tends to $0$ as $p \rightarrow 0.5$.

\begin{figure}
	\centering
		\includegraphics[width=0.45\textwidth]{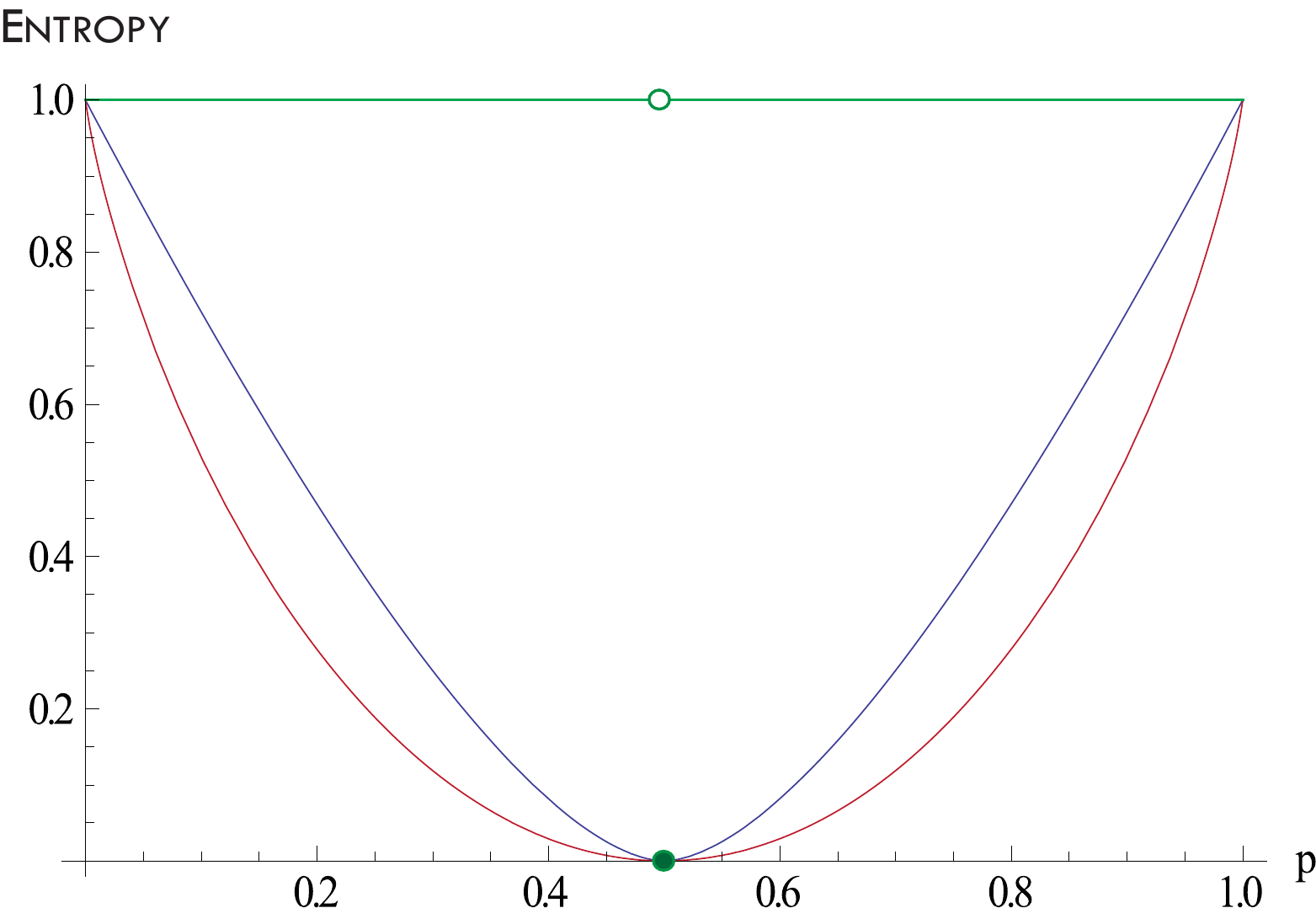}
	\caption{\label{fig:Diagram2}\textbf{Complexity of the Perturbed Coin Simulation.} While the excess entropy of the perturbed coin approaches zero as $p \rightarrow 0.5$ (red line), generating such statistics classically generally requires an entropy of $C_\mu = 1$ (green line). Encoding the past within a quantum system leads to significant improvement (purple line). (Here, $C_q = -\lambda_{+}\log\lambda_{+} -\lambda_{-}\log\lambda_{-}$, where $\lambda_{\pm} = 0.5 \pm \sqrt{p(1-p}$).)  Note, however, that even the quantum protocol still requires an input entropy greater than the excess entropy.}
\end{figure}

Quantum $\epsilon$-machines offer dramatic improvement. We encode the quantum causal states $\ket{S_0} = \sqrt{1-p} \ket{0} + \sqrt{p} \ket{1}$ or $\ket{S_1} = \sqrt{p} \ket{0} + \sqrt{1-p} \ket{1}$ within a qubit, which results in entropy  $C_q = -\tr{\rho\ln\rho}$, where $\rho = \frac{1}{2}(\ket{S_0}\bra{S_0} + \ket{S_1}\bra{S_1})$. The non-orthogonality of $\ket{S_0}$ and $\ket{S_1}$ ensures that this will always be less than $C_\mu$ \cite{benenti07a}. As $p \rightarrow 0.5$, a quantum $\epsilon$-machines tends to require negligible amount of memory to generate the same statistics compared to its classical counterpart (Fig. \ref{fig:Diagram2}).

This improvement is readily apparent when we model a lattice of $K$ independent perturbed coins, which output a number $x\in\mathbb{Z}^{2^K}$ that represents state of the lattice after each perturbation. Any classical model must necessarily differentiate between $2^K$ equally likely causal states, and thus require an input of entropy $K$. A quantum $\epsilon$-machine reduces this to $K C_q$. For $p > 0.2$, $C_q < 0.5$, the initial condition of two perturbed coins may be encoded within a system of entropy $1$. For $p > 0.4$, $C_q < 0.1$; a system of coinciding entropy can simulate $10$ such coins. This indicates that quantum systems can potentially simulate $N$ such coins upon receipt of $K \ll N$ qubits, provided appropriate compression (through lossless encodings~\cite{Bostroem02}) of the relevant past.


\section{Discussion}
In this article, we have demonstrated that any stochastic process with no reversible classical model can be further simplified by quantum processing. Such stochastic processes are almost ubiquitous. Even the statistics of perturbed coins can be simulated by a quantum system of reduced entropy. In addition, the quantum reconstruction can be remarkably simple. Quantum operations on a single qubit, for example, allows construction of a quantum epsilon machine that simulates such perturbed coins. This allows potential for experimental validation with present day technology.

This result has significant implications. Stochastic processes play an ubiquitous role in the modeling of dynamical systems that permeate quantitative science, from climate fluctuations to chemical reaction processes. Classically, the statistical complexity $C_\mu$ is employed as a measure of how much structure a given process exhibits. The rationale is that the optimal simulator of such a process requires at least this much memory. The fact that this memory can be reduced quantum mechanically implies the counterintuitive conclusion that quantizing such simulators can reduce their complexity beyond this classical bound, even if the process they're simulating is purely classical. Many organisms and devices operate based on the ability to predict and thus react to the environment around them. The possibility of exploiting quantum dynamics to make identical predictions with less memory implies that such systems need not be as complex as one originally thought.

This leads to the open question, is it always possible to find an ideal simulator? Certainly, Fig. \ref{fig:Diagram2} shows that our construction, while superior to any classical alternative, is still not wholly reversible. While this irreversibility may indicate that more efficient quantum models exist, it is also possible that ideal models remain forbidden within quantum theory. Both cases are interesting. The former would indicate that the notion of stochastic processes `hiding' information from the present~\cite{crutch09} is merely a construct of inefficient classical probabilistic models, while the latter hints at a source of temporal asymmetry within the framework of quantum mechanics; that it is fundamentally impossible to simulate certain observable statistics reversibly.

\section{Methods:}

\small
\textbf{Proof of Theorem 1.} Let the aforementioned $\epsilon$-machine have causal states $\mathcal{S} = \{S_i\}_1^N$ and emission alphabet $\Sigma$. Consider an instance of the $\epsilon$-machine at a particular time-step $t$. Let $\mathbf{S}_t$ and $\mathbf{X}_t$ be the random variables that respectively governs its causal state and observed output at time $t$, such that the transition probabilities that define the $\epsilon$-machine can be expressed as
\begin{equation}
T_{j,k}^{(r)} = P(\mathbf{S}_t = S_k, \mathbf{X}_t = r|\mathbf{S}_{t-1} = S_j).
\end{equation}
We say an ordered pair $(S_j \in \mathcal{S}, r \in \Sigma)$ is a valid \emph{emission configuration} iff $T_{j,k}^{(r)} \neq 0$ for some $S_k \in \mathcal{S}$. That is, it is possible for an $\epsilon$-machine in state $S_j$ to emit $r$ and transit to some $S_k$. Denote the set of all valid emission configurations by $\Omega_E$. Similarly, we say an ordered pair $(S_k \in \mathcal{S}, r \in \Sigma)$ is a valid reception configuration iff $T_{j,k}^{(r)} \neq 0$ for some $S_j \in \mathcal{S}$, and denote the set of all valid reception configurations by $\Omega_R$.

We define the transition function $f: \Omega_E \rightarrow \Omega_R$. Such that $f(S_j, r ) = (S_k, r ) $ if the $\epsilon$-machine set to state $S_j$ will transition to state $S_k$ upon emission of $r$. We also introduce the shorthand $\mathbf{X}_a^b$ to denote the the list of random variables $\mathbf{X}_a, \mathbf{X}_{a+1},\ldots,\mathbf{X}_b$.

We first prove the following observations.

\begin{enumerate}
\item $f$ is one-to-one iff there exist no distinct causal states, $S_j$ and $S_k$, such that $T_{j,l}^{(r)}, T_{k,l}^{(r)} \neq 0$ for some $S_l$.

    \textbf{Proof:} Suppose $f$ is one-to-one, then $f(S_j, r) = f(S_k, r)$ iff $S_j = S_k$. Thus, there does not exist two distinct causal states, $S_j$ and $S_k$ such that $T_{j,l}^{(r)}, T_{k,l}^{(r)} \neq 0$ for some $S_l$. Conversely, if $f$ is not one-to-one, so that $f(S_j, r) = f(S_k, r)$ for some $S_j \neq S_k$. Let $S_l$ be the state such that $f(S_j, r) = (S_l, r)$, then $T_{j,l}^{(r)}, T_{k,l}^{(r)} \neq 0$.

\item $H(\mathbf{S}_{t-1}|\mathbf{X}_t\mathbf{S}_t) = 0$ iff $f$ is one-to-one.

    \textbf{Proof:} Suppose $f$ is one-to-one. Then for each $(S_j, r) \in \Omega_R$, there exists a unique $(S_k, r)$ such that $f(S_k,r) = (S_j, r)$. Thus, given $\mathbf{S}_t = S_j$ and $\mathbf{X}_t = r$, we may uniquely deduce $S_k$. Therefore $H(\mathbf{S}_{t-1}|\mathbf{X}_t\mathbf{S}_t) = 0$. Conversely, should $H(\mathbf{S}_{t-1}|\mathbf{X}_t\mathbf{S}_t) = 0$, then $H(\mathbf{S}_{t-1}\mathbf{X}_t|\mathbf{X}_t\mathbf{S}_t) = 0$, and thus $f$ is one-to-one.

\item $H(\mathbf{S}_{t-1}|\mathbf{X}_t\mathbf{S}_t) = 0$ implies $H(\mathbf{S}_{t-1}|\mathbf{X}_0^t) =  H(\mathbf{S}_t|\mathbf{X}_0^t)$.

    \textbf{Proof:} Note that (i) $H(\mathbf{S}_{t-1}|\mathbf{X}_0^t\mathbf{S}_t) = H(\mathbf{S}_{t}|\mathbf{X}_{0}^{t}\mathbf{S}_{t-1}) + H(\mathbf{X}_0^t\mathbf{S}_{t-1}) - H(\mathbf{X}_0^t\mathbf{S}_t)$ and (ii) that, since the output of $f$ is unique for a given $(r,S) \in \Omega_E$,    $H(\mathbf{S}_t|\mathbf{X}_{t}\mathbf{S}_{t-1}) = 0$. (ii) implies that $H(\mathbf{S}_{t}|\mathbf{X}_{0}^{t}\mathbf{S}_{t-1}) = 0$ since uncertainty can only decrease with additional knowledge and is bounded below by $0$. Substituting this into (i) results in the relation $H(\mathbf{S}_{t-1}|\mathbf{X}_t\mathbf{S}_t) = H(\mathbf{X}_0^t\mathbf{S}_{t-1}) - H(\mathbf{X}_0^t\mathbf{S}_t)$.Thus   $H(\mathbf{S}_{t-1}|\mathbf{X}_t\mathbf{S}_t) = 0$ implies $H(\mathbf{S}_{t-1}|\mathbf{X}_0^t) =  H(\mathbf{S}_t|\mathbf{X}_0^t)$.

\item $H(\mathbf{S}_{t-1}|\mathbf{X}_0^t) =  H(\mathbf{S}_t|\mathbf{X}_0^t)$ implies $C_\mu = \mathbf{E}$.

    \textbf{Proof:} The result follows then from two known properties of $\epsilon$-machines, (i) $\lim_{t \rightarrow \infty} H(\mathbf{S}_t|\mathbf{X}_0^t) = 0$ and (ii) $C_\mu - E =  H (\mathbf{S}_{-1}|\mathbf{X}_0^\infty$) \cite{crutch01}. Now assume
    that $H(\mathbf{S}_{t-1}|\mathbf{X}_0^t) =  H(\mathbf{S}_t|\mathbf{X}_0^t)$, recursive substitutions imply that $H(\mathbf{S}_{-1}|\mathbf{X}_0^t) =  H(\mathbf{S}_t|\mathbf{X}_0^t)$. In the limit where $t \rightarrow \infty$, the above equality implies $C_\mu - \mathbf{E} = 0$.

\item $C_\mu = \mathbf{E}$ implies $H(\mathbf{S}_{t-1}|\mathbf{X}_t\mathbf{S}_t) = 0$.

    \textbf{Proof:} Since (i) $C_\mu = \mathbf{E} = H(\mathbf{S}_{-1}|\mathbf{X}_0^\infty) \leq H(\mathbf{S}_{-1}|\mathbf{X}_0^\infty \mathbf{S}_0)$, and (ii) $H(\mathbf{S}_{t-1}|\mathbf{X}_t\mathbf{S}_t) = H(\mathbf{S}_{-1}|\mathbf{X}_0\mathbf{S}_0)$, it suffices to show that $H(\mathbf{S}_{-1}|\mathbf{X}_0^\infty \mathbf{S}_0) = H(\mathbf{S}_{-1}|\mathbf{X}_0\mathbf{S}_0)$.

    Now $H(\mathbf{S}_{-1}|\mathbf{X}_0^\infty \mathbf{S}_0)  =  H(\mathbf{X}_0^\infty \mathbf{S}_{-1}\mathbf{S}_0 ) - H(\mathbf{X}_0^\infty \mathbf{S}_0 )
     =  H(\mathbf{X}_1^\infty|\mathbf{S}_{-1}\mathbf{X}_0 \mathbf{S}_0 ) + H(\mathbf{X}_0 \mathbf{S}_{-1} \mathbf{S}_0)
    -  H(\mathbf{X}_1^\infty |\mathbf{X}_0\mathbf{S}_0) - H (\mathbf{X}_0\mathbf{S}_0)$. But, by the Markov property of causal states, $H(\mathbf{X}_1^\infty|\mathbf{S}_{-1}\mathbf{X}_0 \mathbf{S}_0 ) = H(\mathbf{X}_1^\infty|\mathbf{X}_0 \mathbf{S}_0 )$, thus $H(\mathbf{S}_{-1}|\mathbf{X}_0^\infty \mathbf{S}_0) = H(\mathbf{X}_0 \mathbf{S}_{-1} \mathbf{S}_0) - H (\mathbf{X}_0\mathbf{S}_0) = H(\mathbf{S}_{-1}|\mathbf{X}_0\mathbf{S}_0)$, as required.

\end{enumerate}

Combining (1), (2), (3) and (4), we see that there exists a non-zero probability that two distinct causal states, $S_j$ and $S_k$ such that $T_{j,l}^{(r)}, T_{k,l}^{(r)} \neq 0$ for some $S_l$ only if $C_\mu \neq \mathbf{E}$. Meanwhile (1), (2), and (5) imply that there exists no two distinct causal states, $S_j$ and $S_k$ such that $T_{j,l}^{(r)}, T_{k,l}^{(r)} \neq 0$ for some $S_l$ only if $C_\mu = \mathbf{E}$. Theorem 1 follows.

\textbf{Constant Entropy Prediction Protocol.}
Recall that in the simple prediction protocol, the preparation of the next quantum causal state was based on the result of a measurement in basis $\ket{k}$. Thus, although we can encode the initial conditions of a stochastic process within a system of entropy $C_q$, the decoding process requires an interim system of entropy $C_\mu$. While this protocol establishes that quantum models require less knowledge of the past, quantum systems implementing this specific prediction protocol still need $C_\mu$ bits of memory at some stage during their evolution.

This limitation is unnecessary. In this section, we present a more sophisticated protocol whose implementation has entropy $C_q$ at all points of operation. Consider a quantum $\epsilon$-machine initialized in state $\ket{S_j} = \sum_{k = 1}^n \sum_{r \in \Sigma} \sqrt{T_{jk}^{(r)}}\ket{r}\ket{k}$. We refer the subsystem spanned by $\ket{r}$ as $\mathcal{R}_1$, and the subsystem spanned by $\ket{k}$ as $\mathcal{K}$. To generate correct predictive statistics, we

\begin{figure}
	\centering
		 \includegraphics[width=0.45\textwidth]{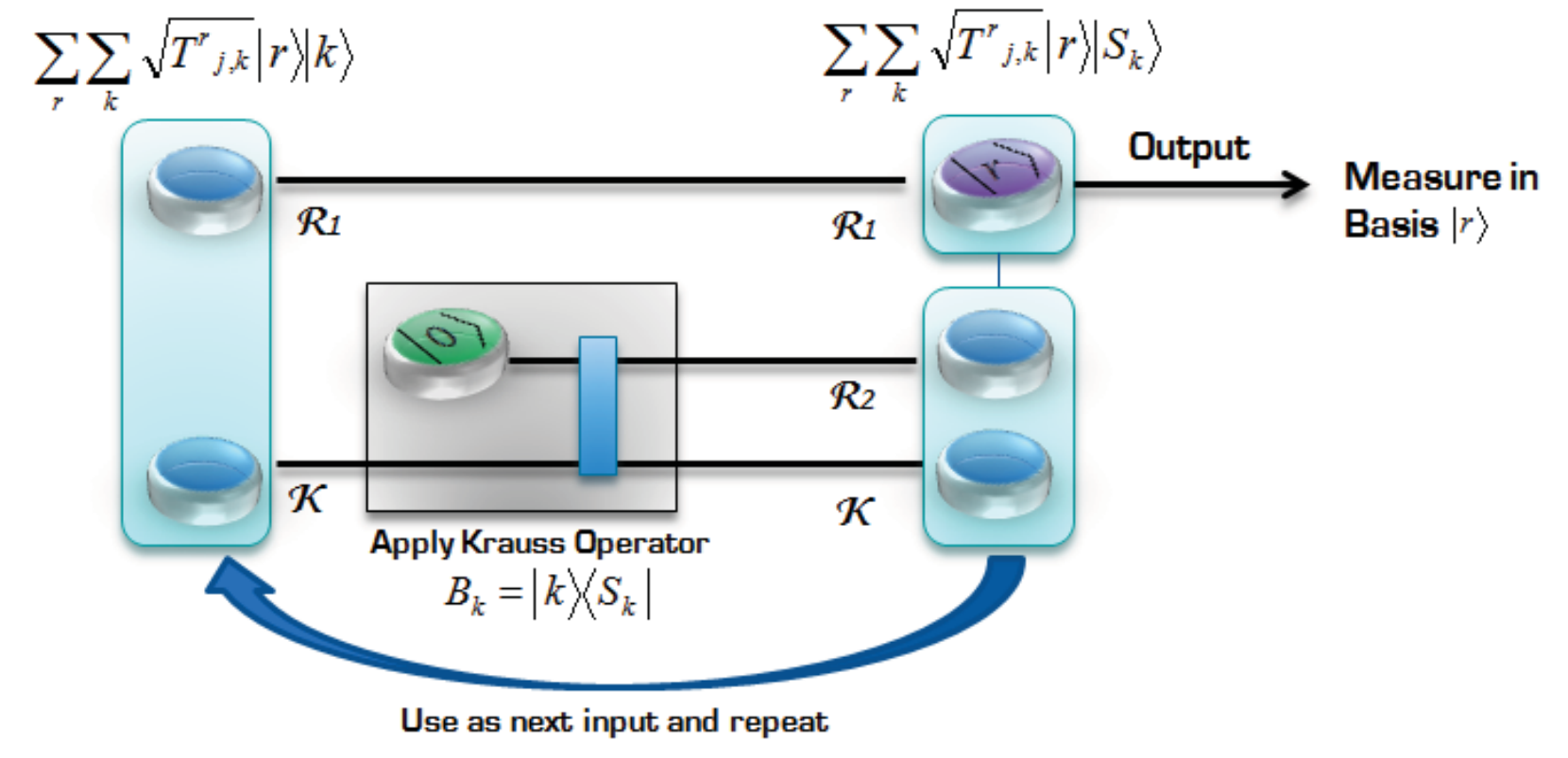}
	\caption{\label{fig:decode_figure}Quantum circuit representation of the refined prediction protocol.}
\end{figure}

\begin{enumerate}
\item Apply a general quantum operation on $\mathcal{K}$ that maps any given $\ket{S_j}$ to $\ket{S'_j} = \sum_{k = 1}^n \sum_{r \in \Sigma} \sqrt{T_{jk}^{(r)}}\ket{r}\ket{S_k}$ on $\mathcal{R}_1\times\mathcal{R}_2\times\mathcal{K}$, where $\mathcal{R}_2$ is a second Hilbert space of dimension $|\Sigma|$. Note that this operation always exists, since it is defined by Krauss operators $B_{k} = \ket{S_k}\bra{k}$ that satisfy $\sum_k B_{k}^\dag B_{k} = 1$.
\item Output $\mathcal{R}_1$. Measurement of $\mathcal{R}_1$ in the $\ket{r}$ basis leads to a classical output $r$ whose statistics coincide with that of its classical counterpart, $x_1$.
\item The remaining subsystem $\mathcal{R}_2\times\mathcal{K}$ is retained as the initial condition of the quantum $\epsilon$-machine at the next timestep.
\end{enumerate}

See Fig. \ref{fig:decode_figure} for a circuit representation of the protocol. Step (1) does not increase system entropy since entropy is conserved under addition of pure ancilla, while $\ip{S'_j}{S'_k} \geq \ip{S_j}{S_k}$ for all $j,k$. Tracing out $\mathcal{R}_1$ in step (3) leaves the epsilon machine in state $\sum p_j \ket{S_j}\ket{S_j}$, which has entropy $C_q$.  Finally, the execution of the protocol does not require knowledge of the measurement result $r$ (In fact, the quantum $\epsilon$-machine can thus execute correctly even if all outputs remained unmeasured, and thus are truly ignorant of which causal state they're in!). Thus, the physical application of the above protocol generates correct predication statistics without requiring more than memory $C_q$.

{\bf Acknowledgments---}  M.G. would like to thank C. Weedbrook, H. Wiseman, M. Hayashi, W. Son and K. Modi for helpful discussions. M.G. and E.R. are supported by the National Research Foundation and Ministry of Education, in Singapore.  K.W. is funded through EPSRC grant EP/E501214/1. V.V. would like to thank EPSRC, QIP IRC, Royal Society and the Wolfson Foundation, National Research Foundation (Singapore) and the Ministry of Education (Singapore) for financial support.

\begin{appendix}

\end{appendix}
\bibliographystyle{nature}
%



\end{document}